\documentclass[aps,pra,twocolumn,superscriptaddress,showpacs,amsmath,amssymb]{revtex4-1}

\usepackage{graphicx}
\usepackage{dcolumn}
\usepackage{bm}

\begin{document}

\title{Formation of Plasmon-Polariton Pulses in the Cooperative Decay of Excitons of Quantum Dots Near a Metal Surface}

\author{A.V. Shesterikov}
\affiliation{Vladimir State University, Vladimir, 600000 Russia}
\author{M.Yu. Gubin}
\affiliation{Vladimir State University, Vladimir, 600000 Russia}
\author{M.G. Gladush}
\affiliation{Institute of Spectroscopy, Russian Academy of Sciences, Troitsk, 142190 Russia}
\author{A.Yu. Leksin}
\affiliation{Vladimir State University, Vladimir, 600000 Russia}
\author{A.V. Prokhorov}
\email{avprokhorov33@mail.ru}
\affiliation{Vladimir State University, Vladimir, 600000 Russia}

\date{\today}

\begin{abstract}
The formation of pulses of surface electromagnetic waves in a metal/dielectric interface is considered
in the process of cooperative decay of excitons of quantum dots distributed near a metal surface in a
dielectric layer. It is shown that the efficiency of exciton energy transfer to excited plasmons can be increased by selecting the dielectric material with specified values of the complex permittivity. It is found
that in the mean field approximation the semiclassical model of formation of plasmon pulses in the system
under study is reduced to the pendulum equation with the additional term of nonlinear losses.
\end{abstract}

\maketitle


\section{\label{sec:1}Introduction}

Collective energy emission processes by a system of
quantum emitters such as optical superradiation have
long been studied, both theoretically and experimentally~\cite{1,2,3,4}. The new possibilities of known effects can
be related to the collective behavior of plasmon oscillators
pumped by the near field of excited chromophores
such as semiconductor quantum dots (QDs),
dye molecules, etc.~\cite{5}. In the case of localized "quantum
dot + metal nanoparticle" systems~\cite{6} or individual
combined core–shell nanocrystals~\cite{7}, their kinematics
is well described by the spaser theory~\cite{5}. However,
plasmons formed in this case are strongly
localized and their collective dynamics is restricted by
the region of action of the near field of plasmon
nanoparticles~\cite{8}.

When a system is extended to the case of a 1-D
array (chain) of localized spasers, the
appearing collective nonlinear regimes can lead to the
considerable narrowing down of emission lines and
the simultaneous compensation of optical losses~\cite{9}.
In the case of 2-D arrays of localized
spasers such as an ensemble of QDs near a metal surface
with defects, there is a region of the collective behavior of
the system caused by the self-synchronization
of individual chromophores due to the near-field
interaction between them~\cite{10}. As a variant, chromophores
can be synchronized by an external pump
beam, which enhances the efficiency of induced processes
in the system under study, resulting in the formation
of a narrow coherent optical beam~\cite{11} perpendicular
to the metal surface. As the external pump, the near field of the tip of a scanning tunneling microscope
can be used~\cite{12}.

Significant interest is the alternative possibility
related to the coherent amplification of the near field of
propagating surface plasmon-polaritons (SPP) due to collective effects
with chromophores under conditions of the partial or
complete suppression of processes of their radiative
relaxation. The problem of the propagation of a plasmon
field appears, in particular, in 1-D
systems such as a metal groove~\cite{13} or a pyramid~\cite{14,15} with nearby QDs and is solved by analyzing Maxwell–
Bloch equations. However, when the decay rate $\gamma_{p}$ of plasmons in a metal is significant, the development
of collective coherent processes involving SPP is much less efficient than emission
processes in optical modes, in particular, in the superradiation
mode.

At the same time, as convenient interfaces for
observing coherent processes with SPP, planar metal/dielectric waveguides already realized in
practice can be used, in which the transverse focusing
of plasmon modes is performed by analogs of Bragg
mirrors~\cite{16}. The solution of the problem of plasmon
decay in such systems can be related to the use of photonic
crystals as a dielectric layer~\cite{17} when long-range SPP are formed in the system with the
field energy maximum considerably shifted to the
dielectric region.

Another way for compensating plasmon decay
in a metal can be the model of a waveguide spaser with
near-field pumping from chromophores located near
a metal surface~\cite{18}. The processes of interaction of
chromophores with the effective field of a plasmon-polariton wave for such a scheme are described in
detail in~\cite{19,20} by the example of solving problems
on the self-induced transparency and formation of
dissipative solitons for plasmon-polariton pulses. The
author of~\cite{21} proposed to realize such a scheme of a
distributed spaser using a dielectric metamaterial film
doped with QDs. However, it is necessary to take into
account that the efficiency of the exciton energy transfer
to a plasmon mode strongly depends on the ratio $r/\lambda_{p}$, where $r$ r is the distance from a chromophore to
the metal surface and $\lambda_{p}$ is the wavelength of a generated plasmon~\cite{22}. At the same time, for $r/\lambda_{p} \ll 1$, the rate $\Gamma_{a}$ of the spontaneous radiative decay of the chromophore
tends to the limiting value $\Gamma_{a}=\left(2/3\right)\Gamma_{0}$, where $\Gamma_{0}$ is the rate of the radiative decay in vacuum. Then,
under the condition $\Gamma_{a} \ll \gamma_{p}$ , the rate of the radiative
decay in a QD could be neglected for this problem.
However, in the presence of a dense (about $10^{15} \; \textrm{cm}^{-3}$ ensemble of adjacent
excited chromophores, $\Gamma_{a}$ can significantly increase~\cite{23,24} in the initial stage of the system evolution due to dipole-dipole interactions. This can
lead to the undesirable transfer of a part of chromophore
energy to optical modes and can initiate the development of cooperative optical effects, including
superradiation~\cite{8,25}. Thus, the partial or complete
suppression of relaxation processes determined by the
radiative decay rate Ãa of excitons in QDs becomes the
additional necessary condition for the observation of
collective processes involving surface plasmons.

In this paper, we propose an approach for selecting
particular chromophores and an appropriate dielectric host-medium to increase the efficiency of energy
transfer from collective excitations of chromophores to SPP modes in a planar metal/dielectric waveguide.
The condition for observing the process is a
considerable decrease in the effective value of $\Gamma_{a}$, which can be caused by local field effects~\cite{26} appearing
upon the disposition of a dense ensemble of chromophores
in a specially selected dielectric matrix. Our
model assumes that the permittivity of the dielectric host-medium is complex, which allows us to completely
compensate the spontaneous relaxation rate of chromophores~\cite{27,28} and to find the qualitatively new
character of their collective dynamics near the metal-dielectric boundary. Discussed collective plasmon-exciton effects can be useful for fast initialization of multiqubits register in plasmonic circuits for quantum computation.

\section{\label{sec:2}Formation of collective SPP generation regimes in a waveguide spaser and basic relations}
\begin{figure*}[t]
\includegraphics[width=1.6\columnwidth]{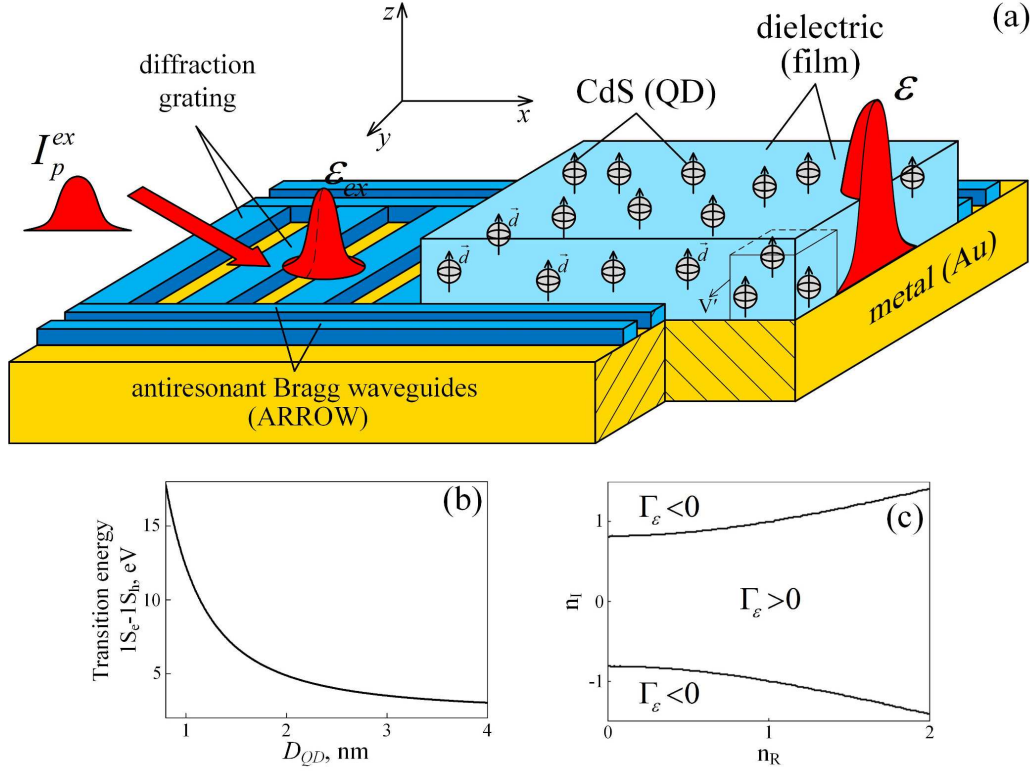}
\caption{\label{fig:1}(a) Formation scheme of SPP pulses in a layered (planar) metal/dielectric waveguide pumped by CdS QDs; (b) dependence of the transition energy on the CdS QD size ($E_{g} =2.42 \; \textrm{eV}$ at $0$~K for a bulk); (c) parametric plane of the complex refractive index $n=n_{R} +in_{I} $ of a dielectric medium with separatrices $\Gamma _{\varepsilon } =0$ for the effective rate of radiative losses of QDs in this medium.}
\end{figure*}

Consider the model of an interface in Fig.~\ref{fig:1}a in the
form of a metal/dielectric waveguide~\cite{29} with two-level
chromophores located inside a thin dielectric
layer, the transition frequency between the two levels $\omega _{a} =2\pi c/\lambda _{a} $ being resonant with the plasmon frequency $\omega _{SPP} =2\pi c/\lambda _{SPP} $ of the metal. By selecting a
dielectric medium with appropriate dispersion characteristics
and providing the initial excitation (inversion)
of a dense ensemble of chromophores in this
model, it is possible to produce the collective decay of
excitons. The difference of this situation from the
model of an emitting spaser~\cite{11} is that the dipole
moments of chromophores are oriented in the direction
perpendicular to the waveguide plane, which
leads to the coherent transfer of their energy predominantly
to SPP modes propagating along the xaxis.
In this case, the process can be localized in the y
direction using the system of additional waveguides
operating based on the Bragg reflection of SPP (antiresonant-reflecting optical waveguide, ARROW~\cite{16}).

By considering the problem in the 3-D
approximation, we assume that the characteristic
size of the interaction region of the effective field of
plasmons and chromophores $h=L_{x} =L_{y} =L_{z} $ satisfies
the inequality $h \ll \lambda _{p} $ and the inequality $L_{z} \ll l_{d} $ is also
valid, where $l_{d} $ is the plasmon decay length along the $z$
axis. In this case, the time-dependent perturbation of
the electron density appearing in the region $V'$ in the
metal causes induced processes in chromophores
located in the symmetric region $V=h^{3} $ in the dielectric.
Then, assuming that SPP modes are quasistatic
within the volume under study~\cite{30}, the corresponding
Rabi frequencies can be written in the form $\Omega =-\left(A\nabla \varphi \mu _{12} a_{0} \right)/\hbar $, where $A=\sqrt{\hbar S / \left(\varepsilon _{0} \varepsilon _{d} \frac{\partial S_{} }{\partial \omega } \right) } $, $a_{0} $ is
the plasmon amplitude, $\mu _{12} $ is the transition dipole
moment in a chromophore, and $\varphi $ is the scalar potential
of the plasmon field linearly decreasing with distance
from the surface, $\hbar$ is the Planck's constant. In the case of excitation of a
mode of the plasmon field at frequency $\omega$, using the
normalization $\int \left|\nabla \varphi \right|^{2} dV=1 $~\cite{30}, the expression for the
Rabi frequency can be approximated by the function $$\Omega =\mu _{12} \sqrt{\frac{S_{n} }{\hbar \varepsilon _{d} \varepsilon _{0} V\frac{\partial S_{n} }{\partial \omega } } } \varepsilon =g\varepsilon,$$ where $\varepsilon =\sqrt{N_{p} } $ and $N_{p} $ is the number of plasmons in
the interaction region.

For a metal-dielectric boundary, the relation $$\lambda _{SPP} =\sqrt{\frac{\textrm{Re}(\varepsilon _{m} )+\textrm{Re}(\varepsilon _{d} )}{\textrm{Re}(\varepsilon _{m} )\textrm{Re}(\varepsilon _{d} )} } \cdot \lambda _{0} $$ is valid, where the parameters $\varepsilon _{d} $ and $\varepsilon _{m} \left(\bar{\omega }\right)=1-\omega _{p}^{2} /\left(\bar{\omega }^{2} +i\gamma _{s} \bar{\omega }\right)$ are the dielectric permittivities of dielectric (with QD) and metal, respectively. Here, $\omega _{p} =\sqrt{4\pi n_{m} e^{2} / m_{0} } $ is the plasma frequency in a
metal, $m_{0} $ and $n_{m} $ are the electron mass and concentration, respectively, $\gamma _{s} $ is the collision
frequency in the metal, $\bar{\omega }=2\pi c/\lambda _{0} $. The spectral properties of the metal-dielectric interface can be described by use of the Bergman's parameter $S\left(\omega \right)=\textrm{Re}\left(\varepsilon _{d} /(\varepsilon _{d} -\varepsilon _{m} \left(\omega \right))\right)$~\cite{5}.

We assume that the pumping volume $V'$ is a
dielectric containing QDs with the characteristic
radius $a$ and concentration $N \gg 10^{21} \; \textrm{m}^{-\textrm{3}}$. The condition $\lambda _{a} \gg a$ allows us to remain within the dipole
approximation, but the large value of dipole transition
moments of QDs~\cite{31} requires the consideration of a
local field acting on emitters~\cite{27,28}. At the same
time, the effects of exchange dipole-dipole interaction~\cite{32} between individual QDs, which are important
in the superradiation problem of a localized spaser~\cite{12}, are neglected in our problem.

Assuming that the refractive index $n=n_{R} +in_{I} $ of
the dielectric environment of QDs is a complex quantity,
where $n=\sqrt{\varepsilon _{d} } $ and $\varepsilon _{d} $ is the complex permittivity,
expressions for the radiative relaxation rate $\Gamma _{a} $, the
Rabi frequency $\Omega $, and the effective frequency detuning $\Delta _{a} $ can be written in the form~\cite{33}
\begin{subequations}
\label{eq:1}
\begin{eqnarray}
\Gamma _{\varepsilon } &=&\Gamma _{a} \left(n_{R} l_{R} -n_{I} l_{I} +2\frac{\delta _{a} }{\Gamma _{a} } l_{I} \right), \\
\Omega _{0} &=&\Omega \cdot \sqrt{l_{R}^{2} +l_{I}^{2} }, \\
\Delta _{\varepsilon } &=&\delta _{a} \left(l_{R} -\frac{\Gamma _{a} }{2\delta _{a} } \left(n_{I} l_{R} +n_{R} l_{I} \right)\right)+\Delta _{a},
\end{eqnarray}
\end{subequations}
where $l\left(n\right)=l_{R} +il_{I} $ is a complex function for which $l_{R} =\left(n_{R}^{2} -n_{I}^{2} \right)/3$, $l_{I} =2n_{R} n_{I} /3$; and $\delta _{a} $ is a small correction
caused by the Lamb shift. It is assumed here that
the function $l\left(n\right)=E_{l} /E_{M} $ coupling the Lorentz local $E_{l} $ and
Maxwell $E_{M} $ fields will retain its structure in the case of
the near field through which plasmons are excited in
the scheme in Fig.~\ref{fig:1}.

The parameter $\Gamma _{a}^{*} =1/\tau _{R} + 1/\tau _{F} $ is the total rate of
radiative (with the time $\tau _{R} =1/\Gamma _{a} $) and nonradiative
(with the time $\tau _{F} $) losses for QDs in vacuum. By using annealing technology~\cite{34}, the time $\tau_{F}$ can
be increased to values comparable to the radiative
time~\cite{35}. At the same time, when a dense (more than $10^{17} \; \textrm{cm}^{-3} $) ensemble of chromophores is located
near the metal boundary, the spontaneous emission
rate can considerably change and, in particular,
increase~\cite{22,24}. Note that the problem of temporal
stability of single QDs during collective energy transfer to SPP in the configuration in Fig.~\ref{fig:1}a
remains open, similarly to the "blinking" problem of
emitting QDs~\cite{36}.

In the semiclassical approximation, the system can
be described similarly to the "metal nanoparticle in a
dielectric with chromophores" spaser model~\cite{5} with
the help of equations for elements of the density matrix $\rho $ of a two-level chromophore:
\begin{subequations}
\label{eq:2}
\begin{eqnarray}
\nonumber
\dot{\rho }_{12} &=&-(i\Delta _{\varepsilon } +\frac{\Gamma _{\varepsilon } }{2} )\rho _{12} \\
&+&\left(i\Omega _{0}^{*} +i\xi _{0} u_{R} \rho _{12} +\xi _{0} u_{I} \rho _{12} \right)n_{21}, \\
\nonumber
\dot{n}_{21} &=&2i\left(\Omega _{0} \rho _{12} -\Omega _{0}^{*} \rho _{21} \right) \\
&-&4\xi _{0} u_{I} \left|\rho _{12} \right|^{2} -\Gamma _{\varepsilon } \left(1+n_{21} \right),
\end{eqnarray}
\end{subequations}
where $\Delta _{a} =2\pi c\left(1/\lambda _{a} -1/\lambda _{SPP} \right)$, $n_{21} =\rho _{22} -\rho _{11} $. The
Rabi frequency can be written as $\Omega _{0} =g\varepsilon \cdot \sqrt{l_{R}^{2} +l_{I}^{2} } $,
where $g=\mu _{12} \sqrt{S_{n} / \left(\hbar \varepsilon _{d} \varepsilon _{0} V\frac{\partial S_{n} }{\partial \omega } \right) } $ is the coupling
constant and $\varepsilon =A_{p} \sqrt{\varepsilon _{d} \varepsilon _{0} V\frac{\partial S_{n} }{\partial \omega } / \left(\hbar S_{n} \right) } $ is the normalized
field with the amplitude $A_{p} $ of the total field
produced by the perturbed electron density in a metal
and the electromagnetic field component in a dielectric.
In the general case, the relation between these
components can be found only during the simultaneous
solution of the evolution equation for the electron
density in a conductor with the specified geometry
and Maxwell's equation~\cite{20}.

The parameter $\xi _{0} =N\mu _{12}^{2} / \left(3\hbar \varepsilon _{0} \right) $ in (\ref{eq:2}) determines
the addition to the Rabi frequency appearing due to
transition from the Maxwell $E_{M} $ to the local field $E_{l} $~\cite{33} acting on a chromophore.

\begin{figure}[t]
\includegraphics[width=\columnwidth]{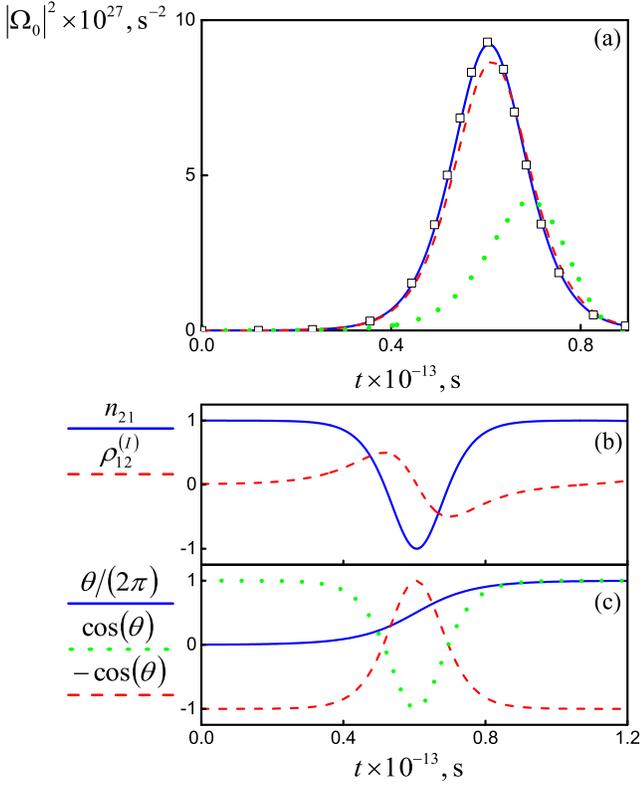}
\caption{\label{fig:2}(a) Profiles of SPP pulse amplitude squared
obtained by the numerical simulation of system (\ref{eq:2})--(\ref{eq:3}) in
the following regimes: (1) neglecting the local field ($\Delta _{\varepsilon } =0$, $u_{I} =0$, $\gamma _{p} =0$, $\Gamma _{\varepsilon } =\Gamma _{a} =6.3 \times 10^{11} \; \textrm{s}^{-1} $ (dashed curve)); (\ref{eq:2}) taking the
local field into account ($\Delta _{\varepsilon } =0$, $u_{I} =-0.1582$, $\gamma _{p} =0$, $\Gamma _{\varepsilon } =0$ (solid curve)); (\ref{eq:3}) for the case $\Delta _{\varepsilon } =0$, $u_{I} =0$, $\gamma _{p} =4.1 \times 10^{13} \; \textrm{s}^{-1} $, $\Gamma _{\varepsilon } =\Gamma _{a} =6.3 \times 10^{11} \; \textrm{s}^{-1} $ (dotted curve). The solution of (\ref{eq:11}) for regime $2$ is shown by squares. (b) Dynamics of parameters $n_{21} $ (solid
curve) and $\rho _{12}^{\left(I\right)} $ (dashed curve). (c) Dynamics of the angle $\theta $ (solid curve) and coefficients $\cos \left(\theta \right)$ (dotted curve) and $-\cos \left(\theta \right)$ (dashed curve) for regime $2$. The interaction parameters are $g=1.37 \times 10^{12} \; \textrm{s}^{-1} $, $\xi _{0} =8.97 \times 10^{11} \; \textrm{s}^{-1} $. The initial polarization of the medium is $\rho _{12}^{} (0)=i\theta _{0} =i/\sqrt{N_{a} } =i \times 1.2 \times 10^{-2} $, the normalization parameter is $\Lambda =9.63 \times 10^{13} \; \textrm{s}^{-1} $ for $N_{a} =7 \times 10^{3} $.}
\end{figure}

The dispersion and dissipative corrections $u_{R} =\left(l_{R} \varepsilon _{R} +l_{I} \varepsilon _{I}\right) / \left(\varepsilon _{R}^{2} +\varepsilon _{I}^{2} \right) $ and $u_{I} =\left(l_{I} \varepsilon _{R} -l_{R} \varepsilon _{I} \right) / \left(\varepsilon _{R}^{2} +\varepsilon _{I}^{2} \right) $,
respectively, are expressed in terms of the real and
imaginary parts of the permittivity of the host-medium~\cite{33} in which QDs are placed and have the
physical meaning of the additional frequency modulation
and the effects of absorption ($u_{I} <0$) or amplification
($u_{I} >0$ ) due to the local field (Fig.~\ref{fig:2}).

To pass to a self-consistent problem, system (\ref{eq:2})
should be supplemented with the equation of motion
for the Rabi frequency of SPP, which in
the case of the exact plasmon resonance has the form
\begin{equation}
\label{eq:3}
\dot{\Omega }_{0} =-\frac{i}{t_{R}^{2} } \rho _{12} -\gamma _{p} \Omega _{0},
\end{equation}
where $$t_{R} =\frac{1}{g\sqrt{N} } =\sqrt{\frac{2\hbar \varepsilon _{d} \varepsilon _{0} \frac{\partial S_{n} }{\partial \omega } }{S_{n} \mu _{12}^{2} N} }$$ determines the characteristic formation time for
quantum correlations in the chosen volume $V'$ in Fig. 1a (compare with the
optical problem~\cite{37} when emitters are located in the
field formation region).

Note that the plasmon mode decay rate $\gamma _{p} =1/\tau _{J} +1/\tau _{R} $ is high and determined by the characteristic times $\tau _{R} $ and $\tau _{J} $ of radiative and "joule" losses, respectively.
Under conditions $1/\tau _{J} \approx 30/\tau _{R} $~\cite{10}, radiative losses
can be neglected, while "joule" losses are determined
by the collision frequency in a metal, i.e., $\gamma _{p} \approx \gamma _{S} $, and in problem (\ref{eq:3}) in the absence of pump , the shortrange SPP appear. In the presence of the
maximum of a surface wave energy in the metal, the
self-consistent problem (\ref{eq:2})--(\ref{eq:3}) will be valid only under conditions
when the characteristic establishment time $t_{R} $ for
correlations between plasmons proves to be considerably
shorter than $\tau _{J} $. Because $t_{R} $ is inversely proportional
to the dipole moment of a chromophore, the
relation $t_{R} <\tau _{J} $ can be valid for pumping a distributed
waveguide spaser by QDs with their giant dipole transition
moments.

In a simple case $\xi _{0} =0$ and in the absence of external
excitation $\varepsilon _{ex} (t)=0$, the collective behavior of the
system, in particular, excitation of superradiation, as
in an optical scheme, is stimulated by a small initial
polarization $\rho _{12}^{\left(I\right)} (0)=\textrm{Im}(\rho _{12}^{} (0))\ne 0$ of the system initiating
the growth of the real part of the Rabi frequency
according to the relation $\dot{\Omega }_{0}^{\left(R\right)} =1/\tau _{R}^{2} \cdot \rho _{12}^{\left(I\right)} $ (see (\ref{eq:3})).
Thus, the front wing of a surface electromagnetic pulse
(SPP pulse) will be formed. Then, according to $\dot{\rho }_{12}^{\left(I\right)} =\Omega _{0}^{\left(R\right)} n_{21} $ from (\ref{eq:2}a) and under the condition $n_{21} >0$, the growth of $\rho _{12}^{\left(I\right)} $ is observed (Fig.~\ref{fig:2}b), which leads
to the excited-level decay according to
$$\dot{n}_{21} =-4\left(\Omega _{0}^{\left(R\right)} \rho _{12}^{\left(I\right)} +\Omega _{0}^{\left(I\right)} \rho _{12}^{\left(R\right)} \right)\approx -4\Omega _{0}^{\left(R\right)} \rho _{12}^{\left(I\right)} $$
from (\ref{eq:2}b). The process will also continue with the
beginning of saturation in the system. However,
because of the change in the sign of the population difference $n_{21} $, the parameter $\rho _{12}^{\left(I\right)} $ gradually decreases to
zero. The condition $\rho _{12}^{\left(I\right)} (t_{D} )=0$ is fulfilled at characteristic
times $t_{D} =t_{R} \ln \left(4/\theta _{0} \right) $ (where $\theta _{0} =1/\sqrt{N_{a} } $ and $N_{a} $ is the number of chromophores in the interaction
region~\cite{38}) when a plasmon pulse is formed. Then this
process is repeated but already in the region where the
parameter $\rho _{12}^{\left(I\right)} $ is negative, which leads to the formation
of the rear edge of the pulse (Fig. 2). A similar picture
is also observed in the case of a low initial stochastic
coherence of the system $\rho _{21}^{\left(R\right)} (0)=Re(\rho _{21}^{} (0))$ in~\cite{39}.

The use of QDs for pumping with their giant dipole
moments at the operating transition can result in a significant
shortening of the establishment time for
quantum correlations and in the proportional decrease
in the delay time $t_{D} $ and duration $t_{W} $ of SPP pulses generated in the system. As a model medium, we use
CdS QDs~\cite{40} located in a dielectric film near the gold
surface. Taking into account the plasmon frequency of
gold $\omega _{p} =1.37 \times 10^{16} \; \textrm{s}^{-1} $ and choosing the condition $\lambda _{0} =387 \; \textrm{nm}$, the wavelength of generated SPP is $\lambda _{SPP} =192 \; \textrm{nm}$. To determine the QD size in the case of the
exact resonance $\Delta _{\varepsilon } =0$, we use the known dependence~\cite{41} of the $1S\left(e\right)\to 1S\left(h\right)$ transition energy on the QD
diameter $D_{QD} =2a$ (Fig.~\ref{fig:1}b)
\begin{equation}
\label{eq:4}
E_{1S(e)-1S(h)} =E_{g} +2\frac{\hbar ^{2} \pi ^{2} }{D_{QD}^{2} } \left(\frac{1}{m_{e} } +\frac{1}{m_{h} } \right)-\frac{3.56\cdot e^{2} }{\varepsilon _{qd} \cdot D_{QD} },
\end{equation}
where $e$ is the electron charge, $m_{e} $ and $m_{h} $ are the effective electron and hole masses,
respectively, in the volume of the QD material with the
permittivity $\varepsilon _{qd} $ and  band gap energy $E_{g}$~\cite{42,43}. The corresponding parameters
for CdS are $m_{e} =0.19m_{0} $, $m_{h} =0.8m_{0} $ and $\varepsilon _{qd} =9$~\cite{32}, which gives $D_{QD} =1.56 \; \textrm{nm}$. Bohr radius of exciton $R_{ex}$ for CdS is $2.5 \; \textrm{nm}$~\cite{Pokutnii} therefore strong confinement regime~\cite{Fedorov} will be observed for the considered QDs, and energy sublevels of conductivity zone will be essentially separated. To tune the QD size
to the plasmon resonance more accurately, it is useful
to employ experimental curves $E_{1S(e)-1S(h)} $ for particular
synthesized QDs~\cite{44}. The dipole moment of the
corresponding interband transition in QDs is assumed
equal to $\mu =\mu _{12} =5 \times 10^{-29} \; \textrm{C} \cdot \textrm{m}$~\cite{5}.

For chosen model parameters and the QD concentration $N = 10^{24} \; \textrm{m}^{-3} $, the characteristic correlation time is $t_{R} =10 \; \textrm{fs}$ and the delay time $t_{D} =60 \; \textrm{fs}$ for the
number of chromophores in the interaction region $N_{a} =7 \times 10^{3} $. The duration of a formed SPP monopulse is only about $14 \; \textrm{fs}$; for regime 1 in Fig.~\ref{fig:2} taking into account the uncompensated rate
of radiative losses, $\Gamma _{\varepsilon } =\Gamma _{a} =6.3 \times 10^{11} \; \textrm{s}^{-1} $ for QDs
near the metal surface~\cite{22}. The additional consideration
of the decay rate of plasmons in gold even under
the condition $\gamma _{p} =4.1 \times 10^{13} \; \textrm{s}^{-1} $ does not strongly
affect on the development of the formation dynamics of
the plasmon pulse (regime 3 in Fig.~\ref{fig:2}a).

However, taking (\ref{eq:1}a) into account, the choice of
the appropriate dielectric host-medium can partially
or completely compensate the increase of $\Gamma _{a} $ (Fig.~\ref{fig:1}c),
but it is also obvious that the properties of natural
media are strongly restricted. Thus, for silica at the
wavelength under study $\lambda _{SPP} =192 \; \textrm{nm}$, we have $n_{R} =1.6$, $n_{I} =5 \times 10^{-7} $~\cite{45} and $\Gamma _{\varepsilon } =2.43\Gamma _{a} $. To completely
compensate relaxation processes in (\ref{eq:2}a) ($\Gamma _{\varepsilon } \equiv 0$), the
required combination of dispersion-dissipative
parameters should satisfy the condition $n_{R} l_{R} -n_{I} l_{I} =0$ (be neglecting a small Lamb shift), which is satisfied,
for example, for the choice $n_{R} =1.6$ and $n_{I} =1.23$.
Such conditions can be fulfilled for an artificial microstructured
dielectric material with specified dispersion–
dissipative characteristics (the Cole-Cole diagram).
They lead to the significant increase in the SPP pulse intensity, while energy transfer from
chromophores to radiation proves to be suppressed
(see regime 2 in Fig.~\ref{fig:2}a). In this case, the influence of
the local field increases, the absolute values of its
parameters increase (corrections $u_{R} =0.37$ and $u_{I} =-0.158$ in (\ref{eq:2})) and the formation dynamics of SPP pulses changes.

\section{\label{sec:3}Collective dynamics of a waveguide spaser in the mean field approximation}
To analyze the contribution of dissipative effects
related to the imaginary part $u_{I} $ of the local field correction,
we can neglect the corresponding phase
effects with $u_{R} $ in (\ref{eq:2}) and decay in (\ref{eq:2})--(\ref{eq:3}) and to pass
in the mean field approximation to a simplified system
of self-consistent equations for a medium
\begin{subequations}
\label{eq:5}
\begin{eqnarray}
\dot{\rho }_{12} &=&\left(i\Omega _{0}^{*} +\xi _{0} u_{I} \rho _{12} \right)n_{21}, \\
\dot{n}_{21} &=&2i\left(\Omega _{0} \rho _{12} -\Omega _{0}^{*} \rho _{21} \right)-4\xi _{0} u_{I} \left|\rho _{12} \right|^{2}
\end{eqnarray}
\end{subequations}
and the effective field
\begin{equation}
\label{eq:6}
\dot{\Omega }_{0} =-ig^{2} N_{a} \rho _{12}
\end{equation}
formed in it.

By passing to the representation for the Rabi frequency
and polarization in the form
$$\Omega _{0} =\frac{1}{2} \left(Ue^{-i\varphi } +U^{*} e^{i\varphi } \right), \; \rho _{12} =\frac{1}{2} R\cdot e^{-iK_{0} }$$
where $K_{0} =\omega _{SPP} t-k_{SPP} z$, and assuming that $Z=n_{21} $, we
can obtain the system of Maxwell-Bloch equations for
a spaser taking into account the (dissipative) local
response of the QD environment
\begin{subequations}
\label{eq:7}
\begin{eqnarray}
\dot{Z}&=&-\frac{i}{2} \left(UR^{*} -U^{*} R\right)-\xi _{0} u_{I} \left|R\right|^{2}, \\
\dot{R}&=&i\left(U-i\xi _{0} u_{I} R\right)Z, \\
\dot{U}&=&-ig^{2} N_{a} R.
\end{eqnarray}
\end{subequations}

The system of equations (\ref{eq:7}) was derived using the
rotating wave approximation by neglecting high-frequency
terms with phase factors $e^{\pm 2i\varphi } $. By passing to
new dimensionless variables $\delta _{0} =-i\frac{U}{\Lambda } $ and $\tau =t\cdot \Lambda $,
where $\Lambda =g\sqrt{N_{a} } $ and setting $R^{*} =R$ and $\delta ^{*} =\delta $, we
represent system (\ref{eq:7}) in the form
\begin{subequations}
\label{eq:8}
\begin{eqnarray}
\frac{\partial Z}{\partial \tau } &=&\delta _{0} R-\frac{\xi _{0} u_{I} }{\Lambda } \left|R\right|^{2}, \\
\frac{\partial R}{\partial \tau } &=&-\delta _{0} Z+\frac{\xi _{0} u_{I} }{\Lambda } RZ, \\
\frac{\partial \delta _{0} }{\partial \tau } &=&-R.
\end{eqnarray}
\end{subequations}
The solution of system (\ref{eq:8}) can be written in the form $Z=B\cos \left(\theta \right)$ and $R=B\sin \left(\theta \right)$, where $B$ and $\theta $ determine
the amplitude and angle of the so-called Bloch
vector with coordinates $Z$ and $R$ and their substitution
to (\ref{eq:8}) gives the equation for the angle
\begin{equation}
\label{eq:9}
\dot{\theta }=-\delta _{0} +\frac{\xi _{0} u_{I} }{\Lambda } B\sin \left(\theta \right).
\end{equation}

By substituting the expression for $\delta _{0} $ from (\ref{eq:9}) into
(\ref{eq:8}c), we obtain a new variant of the pendulum equation
with the nonlinear harmonic losses/decay term
\begin{equation}
\label{eq:10}
\ddot{\theta }-\frac{\xi _{0} u_{I} }{\Lambda } B\cos \left(\theta \right)\cdot \dot{\theta }=B\sin \left(\theta \right).
\end{equation}
The second term in the left-hand side of (\ref{eq:10}) is
responsible for processes initiated by the local
response of the medium and synchronized with the
change in the angle $\theta $. By using the separatrix condition $B=1$ corresponding to the passage from the rotational
motion of the pendulum to vibrational, Eq. (\ref{eq:10})
can be written in the form
\begin{equation}
\label{eq:11}
\ddot{\theta } - K \cos \left(\theta \right)\cdot \dot{\theta }=\sin \left(\theta \right).
\end{equation}
where the amplitude of the decay coefficient is
defined as $K=\xi _{0} u_{I} /\Lambda $. In the absence of the loss modulation,
when $K\cos \left(\theta \right)\cdot \dot{\theta }=K\cdot \dot{\theta }$ , Eq. (\ref{eq:11}) is
reduced to the usual nonlinear pendulum equation
with losses~\cite{38}. Taking the modulation into account
under the same conditions $K<0$ ($\varepsilon _{I} >0$ and $u_{I} <0$), the
pendulum experiences the additional decay in intervals $$\theta \in \left[0+2\pi m;\frac{\pi }{2} +2\pi m\right], \; \theta \in \left[\frac{3\pi }{2} +2\pi m;2\pi +2\pi m\right]$$ responsible for the formation of the leading and trailing edges of SPP pulse (see Fig.~\ref{fig:2}c), whereas in
the interval
$$\theta \in \left[\frac{\pi }{2} +2\pi m;\frac{3\pi }{2} +2\pi m\right]$$
when the central part of SPP pulse is formed,
the enhancement of pendulum oscillations is
observed; $m=0,1,2...$.

In other words, the absorbing dielectric host-medium coherently preserves a part of the QD energy
during the formation of the leading edge of the pulse
and then coherently returns this energy to SPP pulse during formation of the pulse peak. As a result,
taking into account the compensation of the spontaneous
relaxation rate of QDs ($\Gamma _{\varepsilon } =0$) and nonlinear
terms with $u_{I} $ in (\ref{eq:5}), the increase in the peak pulse
intensity is observed with respect to the case when the
response of the host-medium is neglected (see
Fig.~\ref{fig:2}a).

It seems that in the presence of strong resonator
effects in the dielectric host-medium near the QD resonant wavelength, the peak intensity of generated
pulses can be additionally increased due to terms with $u_{R} $ and $u_{I} $ in system (\ref{eq:2}). But because these corrections
are obtained assuming that the spasing wavelength $\lambda _{SPP} $ lies at the wing of the absorption line of a
dielectric film~\cite{33}, this theory cannot be applied to
the given case. Nevertheless, such amplification can
be achieved, for example, using dielectric metamaterials~\cite{46} with a specially selected geometry doped with
QDs~\cite{47,48,49}. In this case, it is possible to excite longrange
surface plasmon polaritons~\cite{17} at a metal/(QDdoped
metamaterial) boundary with their simultaneous
amplification due to QD pumping. However, problem (\ref{eq:2})--(\ref{eq:3}) becomes considerably complicated in this
case, because of the necessity of describing the field
components in a dielectric, the consideration of the
geometry of individual scattering centers~\cite{50} and the
influence of the inhomogeneous structured microenvironment
on the spontaneous decay rate in QDs~\cite{51}.

For the case $u_{I} >0$ ($K>0$)~\cite{33}, when a host-medium with the background amplification exists, on
the contrary, energy transfer from the medium to
emitters doped into it should occur in the initial
stage of the plasmon pulse formation. This causes the
advance of the pulse generation in the medium and a
partial loss in the pulse intensity compared to the case $u_{I} =0$.

In the known case $\xi _{0} =0$, the separatrix solution of
Eq. (\ref{eq:11}) has the form $\theta =4\arctan \left(e^{\tau -\tau _{D} } \right)$, where the
dimensionless delay time is $\tau _{D} =\ln \left(4/\theta _{0} \right) $ with the initial
angle $\theta _{0} =1/\sqrt{N_{a} } $. This corresponds to the formation
of a monopulse with the Rabi frequency modulus
squared that can be written in the form
\begin{equation}
\label{eq:12}
\left|\Omega_{0} \right|^{2} =\frac{\Lambda ^{2} }{\left|\cosh \left(\Lambda \left(t-t_{D} \right)\right)\right|^{2} }
\end{equation}
at the real time scale (regime 1 in Fig.~\ref{fig:2}a).

To simulate Eqs. (\ref{eq:5})--(\ref{eq:11}), we considered a particular
regime assuming that the plasmon phase is $\varphi \left(t\right)=-\pi /2$. In this case, as considered in Section~\ref{sec:2}, the initial
polarization of the system is only imaginary $\rho _{12} \left(0\right)=i R/2$ and the Rabi frequency can be written as $\Omega _{0} =-U^{\left(I\right)} $, where the expansion $U=U^{\left(R\right)} +iU^{\left(I\right)} $ is
used. Under such conditions, the only real part
of the Rabi frequency of the pulse is formed and system
(\ref{eq:8}) is transformed to the system
\begin{subequations}
\label{eq:13}
\begin{eqnarray}
\frac{\partial Z}{\partial \tau } &=&2i\delta _{0} R-\frac{\xi _{0} u_{I} }{\Lambda } \left|R\right|^{2}, \\
\frac{\partial R}{\partial \tau } &=&-2i\delta _{0} Z+\frac{\xi _{0} u_{I} }{\Lambda } RZ, \\
\frac{\partial \delta _{0} }{\partial \tau } &=&\frac{i}{2} R.
\end{eqnarray}
\end{subequations}
However, the form of Eq. (\ref{eq:11}), to which (\ref{eq:13}) is
reduced, does not change under the new normalization
condition $\delta _{0} =-iU^{\left(I\right)} /\Lambda $. The initial conditions in simulation of (\ref{eq:11}) are chosen equal to $\theta _{0} =1/\sqrt{N_{a} } $ for the
initial oscillation angle and
$$\upsilon _{\theta } =\left. \frac{\partial \theta }{\partial t} \right|_{t=0} =\frac{2}{\cosh \left(\ln \frac{\theta _{0} }{4} \right)} $$
for the initial velocity of the pendulum.

Equation (\ref{eq:11}) is a particular case of the Lienard
equation and its approximate analytic solution can be
expressed in terms of elliptic integrals of the first kind.
The numerical solution for the Rabi frequency of SPP pulse field obtained from (\ref{eq:11}) completely coincides with the results of the direct numerical simulation of system (\ref{eq:5})--(\ref{eq:6}) under conditions of the suppression
of spontaneous relaxation in QDs for the
chosen values $n_{R} =1.6$ and $n_{I} =1.23$ ($K=-0.0147$) (see regime 2 in Fig.~\ref{fig:2}a).

\section{\label{sec:4}Influence of concentration and dissipative effects on the development of collective regimes of a waveguide spaser}
\begin{figure}[t]
\includegraphics[width=\columnwidth]{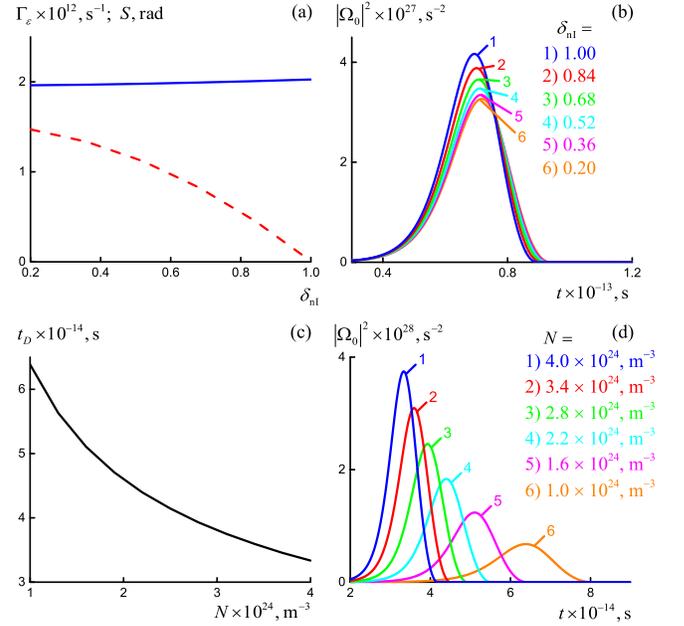}
\caption{\label{fig:3}(a) Dependences of the area $S$ (solid curve) of SPP pulse formed in a waveguide spaser on the relative deviation $\delta _{{\rm nI}} $ of the loss coefficient of as dielectric host-medium from the value $n_{I}^{\sup } =1.23$ for which the suppression of spontaneous relaxation
can be observed for fixed $n_{R} =1.6$; (b) profiles of plasmon pulses for different $\delta _{{\rm nI}} $; (c) Dependence of the pulse delay time $t_{D} $ on the concentration of CdS quantum dots in a dielectric host-medium taking into account local response corrections $u_{I} =-0.1582$ and $u_{R} =0.3754$; (d) pulse profiles at different QD concentrations $N$. The interaction parameters are as in Fig.~\ref{fig:2}; $\Gamma _{a} =6.3 \times 10^{11} \; \textrm{s}^{-1} $.}
\end{figure}

The solutions of Eq. (\ref{eq:11}) are obtained under conditions
of the suppression of spontaneous emission of
excited QDs near a metal/dielectric boundary, whereas
the violation of relations obtained for parameters $n_{R} $ and $n_{I} $ should lead to the increase in the rate of relaxation
processes and the weakening of SPP pulses.
In this case, the efficiency of the QD energy transfer to
the superradition mode nonlinearly depends on the
parameter $\delta _{nI} =n_{I} /n_{I}^{\sup } $ characterizing the relative
deviation of the loss coefficient of the host-medium from the specified level $n_{I}^{\sup } $ for which the condition $\Gamma _{\varepsilon } =0$ is exactly fulfilled for fixed $n_{R} $ (see Fig.~\ref{fig:3}a). In particular,
the decrease in $n_{I}^{} $ leads to the increase in the
relaxation rate $\Gamma _{\varepsilon } $ and the related decrease in the area
$$S=\frac{\mu }{\hbar } \int _{0}^{2t_{D} }A_{p} dt =\int _{0}^{2t_{D} }\left|\Omega _{0} \right|dt \; \left(\textrm{rad}\right)$$
of SPP pulses presented in Fig.~\ref{fig:3}b and calculated
by direct numerical simulation of system (\ref{eq:2})--(\ref{eq:3}).
On the other hand, the change in the QD concentration
under the condition $\Gamma _{\varepsilon } =0$ in the system,
allows one to control the delay time $t_{D} $ and pulse duration $t_{W} $ at the medium output. Thus, the increase in the
QD concentration for CdS leads to the rapid nonlinear
shortening of the delay time $t_{D} $ in the model under
study (Fig.~\ref{fig:3}c) with the emission more and more
intense pulses (Fig.~\ref{fig:3}d).

To analyze nonlinear phase effects during the generation
of plasmon pulses, it is necessary to take into
account the spatial dynamics of collective processes by
introducing the longitudinal coordinate $x$ into Eq. (\ref{eq:3}).
The corresponding solution at the output of a medium
of length $L$ with the isotropic distributions of chromophores
will have the form~\cite{37}
\begin{eqnarray}
\nonumber
\varepsilon '(L,t)&=&\varepsilon _{ex} (t)\cdot \exp (ikL) \\
&+&i\frac{1}{\tau _{R} g\varepsilon _{0} L} \int _{0}^{L}\rho _{21} \cdot \exp (ik\left|L-x'\right|)dx', \label{eq:14}
\end{eqnarray}
where $\tau _{R} =t_{R}^{2} \cdot \frac{c}{L} $ is determined by the new characteristic
formation time of quantum correlations, $\varepsilon _{ex} (t)$ is
the amplitude of an additional trigger SPP pulse at the entrance of waveguide, and the factor $\exp (ik\left|L-x'\right|)$ gives
phase shifts which, unlike the case considered in
Section~\ref{sec:3}, depend on the coordinate $x$.

It is convenient to study the influence of nonlinear
dispersion effects on the spectral features of SPP pulses beginning from the numerical solution of the
joint system of equations (\ref{eq:2}) and (\ref{eq:14}) under conditions $\varepsilon _{ex} (t)=0$ when a trigger pulse at the medium
input is absent and also by neglecting dissipative terms
of the local field $u_{I} =0$.

The real spectral shape of pulses generated in such
approximation is determined by the inverse Fourier transform from the corresponding Rabi frequency at
the medium output
\begin{eqnarray}
\nonumber
F(L,\omega )&=&\left|\Omega _{0} (L,\omega )\right|^{2} \\
&=&\left|\frac{g\cdot \sqrt{l_{R}^{2} +l_{I}^{2} } }{\sqrt{2\pi } } \int _{-\infty }^{\infty }\varepsilon '(L,t) e^{-i\omega t} dt\right|^{2}. \label{eq:15}
\end{eqnarray}

In a simple case in the absence of the frequency
modulation and neglecting delay effects, the full width
at half maximum $\Delta \nu _{1/2} $ of a spectrally limited pulse is
determined only by its duration $\tau _{1/2} $. For a pulse in the
form of a hyperbolic secant (\ref{eq:12}), the relation $\Delta \nu _{1/2} \tau _{1/2} =K_{0} $ is valid, where $K_{0} =0.315$~\cite{52}.

\begin{figure}[t]
\includegraphics[width=\columnwidth]{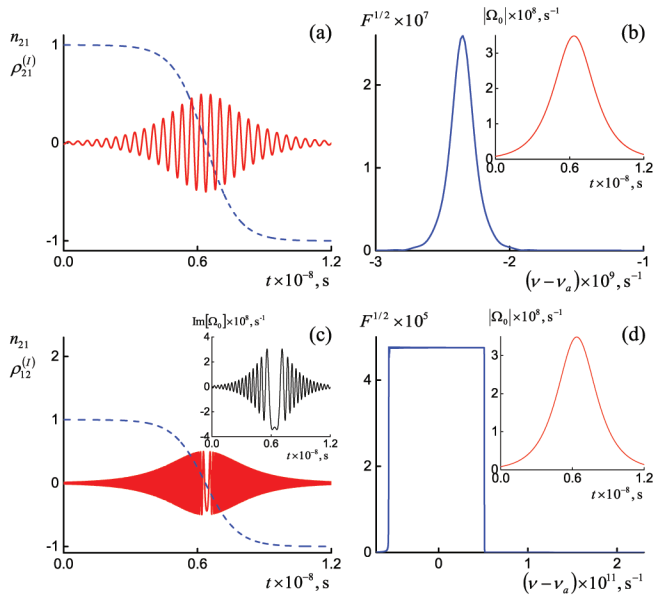}
\caption{\label{fig:4}Time dependences for the population difference $n_{21} $ (dashed curves) and the imaginary part of polarization $\rho _{12}^{(I)} $ (solid curves) in the collective formation regime of SPP on a gold surface initiated by the decay of excitons in a dense ensemble of CdS quantum dots in a dielectric matrix near the metal-dielectric interface under condition $\Delta _{\varepsilon } =-1.48 \times 10^{10} \; \textrm{s}^{-1} $ ($\Delta _{a} =1.04 \times 10^{12} \; \textrm{s}^{-1} $) by neglecting ($u_{R} =0$) (a) and taking into account (c) nonlinear local-field dispersion effects for $u_{R} =0.375$, and also the corresponding
frequency spectra (b, d) calculated by (\ref{eq:15}). The insets in Figs.~\ref{fig:4}b,d show the profiles of SPP pulses and Fig.~\ref{fig:4}c --- the frequency modulation shape $\Omega_{0}^{\left(I\right)} $ of the pulse in the regime with $\Delta _{\varepsilon } =0$ and $u_{R} =0.02$. The interaction parameters are as in Fig.~\ref{fig:2}.}
\end{figure}

Figures~\ref{fig:4}a,b show that the spontaneous relaxation $\Gamma _{\varepsilon } \equiv 0$ is suppressed
rate ($n_{R} =1.6$, $n_{I} =1.23$) but the
effective frequency detuning $\Delta _{\varepsilon } $ is simultaneously
formed in the system (see (\ref{eq:1}c)) corresponding to the
appearance of linear dispersion. It leads, according to
(\ref{eq:14}), to the formation of dispersion delays and the
appearance of characteristic oscillations of the parameter $\rho _{21}^{\left(I\right)} $ (and also of $\textrm{Im}(\Omega _{0} )$) at frequency $\Delta _{\varepsilon } $. The relation
between the spectral width $\Delta \nu _{1/2}^{SR} $ and duration $\tau _{1/2}^{SR} $ of such a modulated pulse changes to $\Delta \nu _{1/2}^{SR} \tau _{1/2}^{SR} =K_{SR} $,
where $K_{SR} =0.8$ (see Figs.\ref{fig:4}a,~\ref{fig:4}b). However, nonlinear phase effects corresponding to the contribute of terms
with $u_{R} $ into (\ref{eq:2}a) were neglected in Figs.~\ref{fig:4}a,~\ref{fig:4}b.

On the other hand, the consideration of only nonlinear
terms $\xi _{0} u_{R} \rho _{21} \left(\rho _{22} -\rho _{11} \right)$ in (\ref{eq:2}a) in the absence of linear dispersion ($\Delta _{\varepsilon } =0$) leads to a strong nonlinear frequency
modulation of the parameter $\rho _{21}^{\left(I\right)} $, which, according to
(\ref{eq:14}) is superimposed on the profile of a generated plasmon pulse. The specific feature of such a modulation
is manifested in the change of its sign with displacement
from the wing of SPP pulse, where $n_{21} =1$, to its peak, where $n_{21} =-1$, and in the formation
of the characteristic profile of the imaginary component of the Rabi frequency (see the inset in Fig.~\ref{eq:4}c
for arbitrarily chosen $u_{R} =0.02$).

Under model conditions, for the chosen values $n_{R} =1.6$ and $n_{I} =1.23$, the calculated value of $u_{R} $ will
be $0.375$ and the effects of linear and nonlinear dispersion
will act simultaneously. As a result, a mixed
regime with the phase modulation rate nonlinearly
increasing from the pulse front to its tail appears in the
system (see Fig.~\ref{fig:4}c). The relation between the duration $\tau _{1/2}^{FM} $ of such a modulated pulse and its spectral width $\Delta \nu _{1/2}^{FM} $ takes the form $\Delta \nu _{1/2}^{FM} \tau _{1/2}^{FM} =K_{FM} $, where $K_{FM} =440$, and its spectrum significantly broadens, becoming
in fact rectangular (Fig.~\ref{fig:4}d). This result obtained
for a distributed waveguide spaser in a pulsed regime
considerably differs from the case of localized spaser
with the characteristic spectral narrowing effect~\cite{53}.
The spectral broadening regime for SPP pulse
for the interface presented in Fig.~\ref{fig:1} can find new
applications in the problems of the development of
broadband electromagnetic sources~\cite{54}, similarly to
the generation of laser combs in optics~\cite{55}.

\section{\label{sec:5}Features of the triggered regime of a waveguide spaser}
The feature of the triggered regime in the scheme in
Fig.~\ref{fig:1}a, similarly to triggered optical superradiation
(TSR), is related to the possibility of controlling the
development of cooperative process in a system when
the establishment of quantum correlations between
individual chromophores is initiated by the external
pump pulse. In this case, the development dynamics,
the radiation pattern and the shape of a supperradiation
pulse are completely determined by the parameters
of this trigger pulse. In optics, such a regime was
first observed in gas medium in~\cite{56}. However, only the
realization of this effect in solids~\cite{57} provided the
basis for using TSR for the development of optical
memory and optical computing devices~\cite{58}. The
translation of this problem to plasmonics offers a
number of advantages, retaining, on the one hand,
optical data processing rates and, on the other hand,
considerably simplifying the integration of individual plasmonic devices in circuits and providing their coupling with electronic computing devices.

In the problem (\ref{eq:2})--(\ref{eq:4}), the triggered regime of generation of
plasmon pulses can be achieved in the presence of a trigger SPP pulse
\begin{equation}
\label{eq:16}
\varepsilon _{ex} (t)=\varepsilon _{0} e^{-\left(t-nT_{0} \right)^{2} /\left(2T_{0}^{2} \right)}
\end{equation}
with duration $T_{0} $ and time delay $nT_{0} $ ($n\in R^{+} $) with
respect to the beginning of the free evolution of the
system due to a relaxation process (see Section~\ref{sec:2}). The
trigger pulse can be obtained by transforming an external
optical pulse on a metal grating, as in~\cite{59}
(Fig.~\ref{fig:1}a). As in the optical case, the specific feature of
the regime is the possibility of controlling the delay time of the main SPP pulse~\cite{60}, which in the
classical formulation of the problem by neglecting
local field effects is determined by the expression
\begin{equation}
\label{eq:17}
\tau {}_{D} =\tau {}_{R} \ln \left(\frac{1+\cos \theta }{1-\cos \theta } \right)
\end{equation}
and depends on the trigger pulse area
$$\theta =\frac{2\mu }{\hbar } \int A_{p}^{ex} dt.$$
In this case, the amplitude $A_{p}^{ex} $ of the optical trigger
pulse can be recalculated to parameters (\ref{eq:16}) according
to the relation
$$\varepsilon _{ex} =A_{p}^{ex} \sqrt{\frac{\varepsilon _{d} \varepsilon _{0} V\frac{\partial S_{n} }{\partial \bar{\omega }} }{\hbar S_{n} } } $$
for the ideal case when $100 \%$ of the optical pulse
energy transfer to a surface wave.

\begin{figure}[t]
\includegraphics[width=\columnwidth]{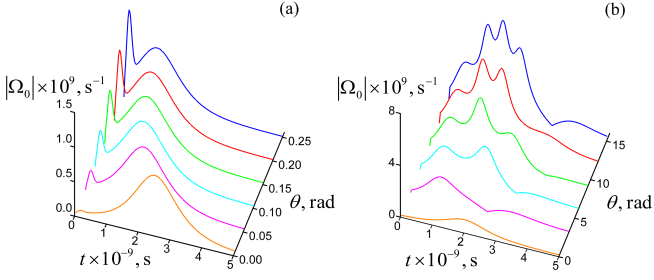}
\caption{\label{fig:5} Profiles of a trigger SPP pulse and a following main pulse as functions of time for different areas $\theta $ of the trigger pulse with the delay coefficient $n=2$ in a host-medium with length $L=\lambda _{a} $ with CdS quantum dots at concentration $N=10^{24} \; \textrm{m}^{-3} $. The simulation parameters are as in Fig.~\ref{fig:2}; (a) the trigger pulse duration is $T_{0} =80 \; \textrm{ps}$, the range of its peak intensities is $I_{0}^{ex} \in \left(1;1.5 \times 10^{3} \right) \; \textrm{W}/\textrm{m}^{2} $; (b) $T_{0} =800 \; \textrm{ps}$, $I_{0}^{ex} \in \left(10;5.5 \times 10^{4} \right) \; \textrm{W}/\textrm{m}^{2} $.}
\end{figure}

Figure~\ref{fig:5}a presents the results of simulating system
(\ref{eq:2}) + (\ref{eq:14}) in the form of a set of the time dependences
of the Rabi frequencies of main pulses produced under
the action of input trigger pulses with different areas
with increasing their peak intensity $I_{0}^{ex} =\left(A_{p0}^{ex} \right)^{2} C_{0} $, where $C_{0} =c\varepsilon _{0} /\left(2n_{R} \right) $. The corresponding dependences
for the delay times of SPP pulse formation are also
approximated by expression (\ref{eq:17}) taking into account
that recalculation expressions between the normalized
time $\tau $ and the real time $t$ are analogous to the passage
from system (\ref{eq:7}) to (\ref{eq:8}). For the chosen combination of
the QD concentration, the duration and power of trigger
pulses, the profile of the main formed pulse in
Fig.~\ref{fig:5}a remains virtually invariable.

\begin{figure}[t]
\includegraphics[width=\columnwidth]{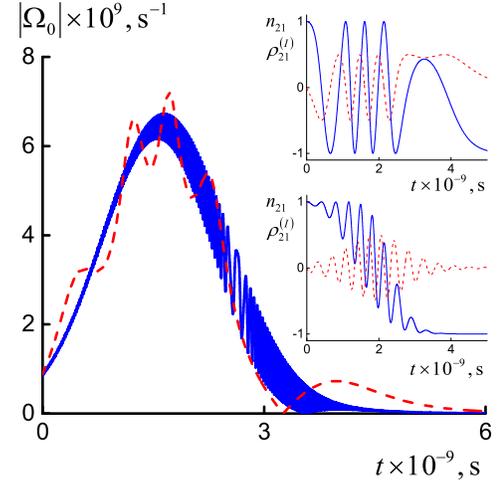}
\caption{\label{fig:6}The temporal profile of SPP pulse for triggered regime in metal/dielectric waveguide spaser with CdS QDs without accounting dispersion corrections of the local field (dashed curves) and with accounting it $u_{R} =0.3754$ (solid curves) ($u_{I} =0$). The parameters correspond to Fig.~\ref{fig:5} provided that the trigger pulse (\ref{eq:16}) is acting with $I_{0}^{ex} =5.5 \times 10^{4} \; \textrm{W}/\textrm{m}^{2} $ (with area $\theta =15.2 \; \textrm{rad}$), $T_{0} =800 \; \textrm{ps}$ and delay coefficient $n=2$ under conditions $\Gamma _{\varepsilon } =0$ and $\Delta _{\varepsilon } =0$. The insets show: time dependences of the polarization component $\rho _{21}^{\left(I\right)} $ (dashed curves) and population difference $n_{21} $ (solid curves) in the absence of frequency shift $\Delta _{a} =0$ (upper inset) and with it $\Delta _{\varepsilon } =-1.48 \times 10^{10} \; \textrm{s}^{-1} $ ($\Delta _{a} =1.04 \times 10^{12} \; \textrm{s}^{-1} $) (lower inset).}
\end{figure}

The regime will qualitatively change when the trigger
pulse duration (\ref{eq:16}) becomes close to the characteristic
duration $t_{W} $ of the main SPP pulse (transition
regime) and its delay $nT_{0} $ is selected so that its envelope
partially or completely overlaps the envelope of the generated pulse. Under such conditions, the field
intensity at the medium output can exhibit a multipeak
structure (see Figs.~\ref{fig:5}b and~\ref{fig:6}).

A similar superradiation regime is well known in
optics and is caused by the nonmonotonic
decay of the excited state of continuous media~\cite{61}.
However, in the case under study for $\Gamma _{\varepsilon } =0$ and $\Delta _{\varepsilon } =0$, the effect is caused by modulation instabilities in the process of QDs excitons decay resulting in the appearance of Rabi oscillations (see
the upper inset in Fig.~\ref{fig:6}) neglecting dispersion effects
with $u_{R} $.

When nonzero frequency detunings $\Delta _{\varepsilon } $ are taken
into account in system (\ref{eq:2}), the time synchronization
of oscillations of the polarization component $\rho _{21}^{\left(I\right)} $ and
the population difference $n_{21} $ is violated. As a result,
the amplitude of oscillations appearing in the system is
modulated by a decreasing function of time proportional to the inverse detuning frequency $\Delta _{\varepsilon } $ (see the
lower inset in Fig.~\ref{fig:6}) in the approximation $u_{R} =0$.

The consideration of the influence of a dielectric host-medium with the dispersion coefficient $u_{R} $ again leads to a strong nonlinear frequency modulation
of the produced pulse (similarly to Figs.~\ref{eq:4}c,~\ref{eq:4}d).
However, in the case of its interference with trigger
pulse (\ref{eq:16}) with appropriate duration (as in Fig.~\ref{fig:5}b),
the envelope of the resulting pulse acquires a strong
high-frequency amplitude modulation, which is
absent for $u_{R} =0$ in Fig.~\ref{fig:6}. The spectra and envelope shape of
the SPP pulse can be recorded by performing
the inverse transformation of surface waves to
an optical signal on a metal grating [29]. Note that a
noticeable change in the permittivity of a metal due to
optical excitation of electrons is observed at the energy
density on the order of $0.5 \; \textrm{mJ}/\textrm{cm}^{2}$~\cite{62}. This allows
one to realize "pure" plasmon nonlinearities~\cite{29} and
perform direct signal-pump experiments with surface
plasmon-polaritons. However, after conversion to
dimensional parameters, the energy density of emitted
pulses in Fig.~\ref{fig:5} does not exceed $0.2 \times 10^{-4} \; \textrm{mJ}/\textrm{cm}^{2}$ and
therefore conditions for these nonlinear regimes are
not achieved in this work.

It is necessary to note, that the contribution of dissipative effects
of the local field to problem (\ref{eq:2}) can be estimated as
$$\delta _{loc}^{\left(I\right)} =\frac{\xi _{0} u_{I} \rho _{21}^{(R)} }{\Omega _{0}^{I} } =\frac{2}{3} \frac{cu_{I} }{\omega _{a} I^{(R)} }, $$
where
$$I^{(R)} =\int _{0}^{L}\exp (ik_{a} \left|L-x'\right|)dx' $$
and for $L\approx \lambda _{a} $, we have $\delta _{loc}^{\left(I\right)} \approx \frac{2}{3} u_{I} $. Similarly, we can
obtain the estimate for the dispersion coefficient
$$\delta _{loc}^{\left(R\right)} =\frac{\xi _{0} u_{R} \rho _{21}^{\left(R\right)} }{\Omega _{0}^{R} } \approx \frac{2}{3} u_{R}, $$
determining the relative contribution of local field
effects to the frequency modulation of the produced
signal. Thus, the contribution of local field effects to
the kinematics of the system under study depends only
on the introduced coefficients $u_{I} $ and $u_{R} $ determined only by the material parameters of the host-medium, but not by its geometry.

\section{\label{sec:6}Conclusions}
We have proposed new efficient methods for the
formation and external control of short SPP pulses at the interface of a metal and a QD-doped
dielectric medium.
The conditions for selecting parameters of QDs and a
dielectric host-medium are determined which provide
the maximal collective energy transfer from a QD
ensemble to SPP modes dominating over the radiative
relaxation of individual chromophores. By the
example of a model medium with CdS nanocrystals,
the dimensional and concentration dependences of
the effect are studied and the amplitude and spectral
features of SPP pulses generated in the system are determined. The presented model and studied regimes can be used, in particular, for solving a practical problem
of increasing the characteristic coherent lengths of
the SPP field.

Our approach can be realized in experiments by
using dielectric films doped with semiconductor QDs
with diameters selected to provide the equality of energies
of interband transition and plasmons excited at the
metal-dielectric interface. However, it is necessary to
take into account that the efficiency of energy transfer
from excitons to plasmon modes can be affected by
blinking, as in the case of luminescent QDs~\cite{36}. In
addition, the physical characteristics of QDs significantly differ from perfect and strongly depend on the
method of their synthesis and characteristics of the host-medium~\cite{63}. In this case, the use of organic
molecules can serve as an alternative for interface
pumping (Fig.~\ref{fig:1})~\cite{64}.

The models presented in the paper can be useful for
practical applications such as the development of plsamonic integrated circuits for quantum computations. In particular, considerated collective effects can be used as a basis for  multiqubits register initialization in the process of formation the quantum correlations between QD. The advantage of the realization of such a register in the plasmon-exciton systems to atomic-optical systems is the ability to implement an effective addressing schemes by coupling of each quantum dot with localized plasmon modes on the nanoscale. However, this requires complication of the circuit shown in Fig.~\ref{fig:1}.
Besides, important problems of the direct connection of such systems with all-optical data communication
systems remain open. In particular, one of the
problems is increasing the efficiency of mutually
reversible conversion of the light wave field and plasmon–
polaritons formed in layered structures~\cite{65}.
Final answers to these problems can be obtained in
relevant experiments, in particular, using epiluminescence
spectromicroscopy of single quantum emitters~\cite{66,67,68}.

Another important technical problem is achieving
very high QD concentrations in a matrix which for the
maximum concentration $N=4 \times 10^{24} \; \textrm{m}^{-3} $ used in this
paper (Fig.~\ref{fig:3}c) will amount to $1.5 \%$ of the concentration $N_{a}^{D} =D_{QD}^{-3} $ of the closest packing of QDs with
diameter $D_{QD} =1.56 \; \textrm{nm}$. One of the solutions can be
using the self-organization of QDs with different sizes
during their evaporation from colloid solutions~\cite{69} on
a substrate. However, the prospects for using such
structures under conditions of the problem under
study require special studies due to a considerable dispersion
of QDs in size.

Note in conclusion that it is also important to
obtain a more general nonlinear equation describing
the propagation of ultrashort SPP
pulses in experiments taking into account nonstationary
terms of the nonlinear dispersion type, etc.~\cite{20,70}. Such terms can appear due to modification of the
metal permittivity by high-power pump femtosecond
pulses of an external optical pump~\cite{29} and due to nonlinear effects in semiconductor
QDs~\cite{71} and in a dielectric host-medium containing them~\cite{72}. Such an equation can
serve as a starting point for searching and determining
the stability conditions~\cite{73} for its soliton solutions
and the development of new schemes of active
nanoplasmonics~\cite{74} with dissipative SPP solitons.

\section{Acknowledgments}
One of the authors (A.V.P) thanks A.B. Evlukhin
for useful discussions. The work was supported by the
Russian Foundation for Basic Research (project nos. 14-02-97511, 14-29-07270 ofi\underline{ }m) and the Ministry of
Education and Science of the Russian Federation
(task VLSU no. 2014/13).


\bibliography{Shesterikov}

\end{document}